\documentclass[conference]{IEEEtran}
\IEEEoverridecommandlockouts
\usepackage{cite}
\usepackage{amsmath,amssymb,amsfonts}
\usepackage{algorithmic}
\usepackage{graphicx}
\usepackage{textcomp}
\usepackage{xcolor}
\usepackage{siunitx}
\usepackage{multirow}
\def\BibTeX{{\rm B\kern-.05em{\sc i\kern-.025em b}\kern-.08em
    T\kern-.1667em\lower.7ex\hbox{E}\kern-.125emX}}
\begin{document}

\title{Development of a Power Quality Based Digital Energy Meter Educational Platform}

\author{\IEEEauthorblockN{1\textsuperscript{st} Mislav Bui}
\IEEEauthorblockA{\textit{Faculty of Electrical Engineering and Computing} \\
\textit{University of Zagreb}\\
Zagreb, Croatia \\
mislav.bui@fer.hr}
\and
\IEEEauthorblockN{2\textsuperscript{nd} Marko Jurčević}
\IEEEauthorblockA{\textit{Faculty of Electrical Engineering and Computing} \\
\textit{University of Zagreb}\\
Zagreb, Croatia \\
marko.jurcevic@fer.hr}
}

\maketitle

\begin{abstract}
Phasor Measurement Units (PMUs) are being used extensively for electrical grid monitoring and control. However, their cost prohibits further adoption on the distribution grid and easy access for educational purposes. This paper proposes that simple and fundamental functions of a PMU can be achieved using an energy metering IC and integrated into smart electricity meters, providing a lower cost and more widely available method of monitoring and control of distribution grids, and presents a proof-of-concept platform with aforementioned functionality. The described platform's construction and flexibility emphasizes its educational capabilities PMUs and electricity meters as well.
\end{abstract}


\section{Introduction}
Phasor Measurement Units (PMUs) are devices which have the ability to achieve high-precision measurements of voltage and current, providing synchrophasors - phasor values referenced to specific points in time \cite{synchrophasor}. PMUs can be used in various ways with most the common application of monitoring and controlling electrical grids in real time, increasing understanding of grid instabilities and faults, as well as locating them and protecting the grid from such events. In addition to real-time monitoring, the collected data can be used for post-mortem analyses to prevent further problem occurrences. They have thus become a popular and advanced tool used to monitor the electrical grid, helping to enforce strict stability, reliability and grid protection requirements. However, their high cost is still limiting more widespread adoption and usage, necessitating scarce and highly optimized placement, for which specific algorithms have been developed. Strategic placement can be used to monitor, control and protect specific parts of infrastructure, especially of transmission grids and substations, but lacks flexibility and wide coverage needed for distribution grid monitoring \cite{distr_grid_PMU1}. 

The distribution grid is projected to become more unstable and will need better monitoring and control due to the impact of distributed energy resources (DER) and active loads. Renewable energy sources are becoming more widely available, even to consumers, who are becoming producers as well (or so-called prosumers), extensively using small-scale photovoltaic panels and wind turbines. This means that unlike traditional grids, smart grids of the future will have a large number of sources, many of them connected directly to the distribution grid, not the transmission grid. All of these factors indicate that the distribution grid is becoming increasingly dynamic and large-scale deployment of PMUs will be needed regardless of cost \cite{distr_grid_PMU2}.

In contrast to PMUs, electricity meters are very common devices on the distribution grid. The smart energy meter is novel type of electricity meter also becoming more common in recent years owing to its advanced features such as automated and real-time reading, measurement of additional properties, such as phase angle and frequency in addition to energy consumption, as well as two-way communication. 

This paper describes a simple, low-cost, prototype proof-of-concept PMU-like platform based on a common all-in-one energy metering IC, the Cirrus Logic CS5463. The prototype platform is a device connected to a precise time source with grid voltage and current inputs and can measure many different aspects of input signals. Such a device could be deployed alongside regular PMUs (on grid locations where PMUs are not presently installed) and enhance their functionality, especially for post-mortem analyses of grid-related problems. With further development such functionality could be implemented directly in smart electricity meters, providing analysis of issues in real-time at the level of distribution grids.
Application of machine-learning (ML) or tiny machine-learning algorithms (TinyML) could further advance real-time monitoring and recognition of grid-related events of interest.
Additionally, a platform like this can be applied for educational purposes as well. The spread of PMU usage means that students - future engineers and technicians - will need to be familiarized with synchrophasors and related devices through education and training, both theoretical and practical. This platform could aid in education and training due to its low cost and high flexibility and adaptability, since every part of it is exposed and accessible to the user, who can modify the platform's behavior as well.
\\
\section{Platform description}
The proposed measurement platform consists of the following three subsystems:
\begin{itemize}
    \item front-end and energy measurement IC,
    \item microcontroller (MCU) operating the measuring process and timestamping measured data,
    \item server used for receiving data and providing accurate time to the MCU.
\end{itemize}
A general overview is provided in Fig. 1.
\\
\begin{figure}[htbp]
\centerline{\includegraphics[scale=0.3]{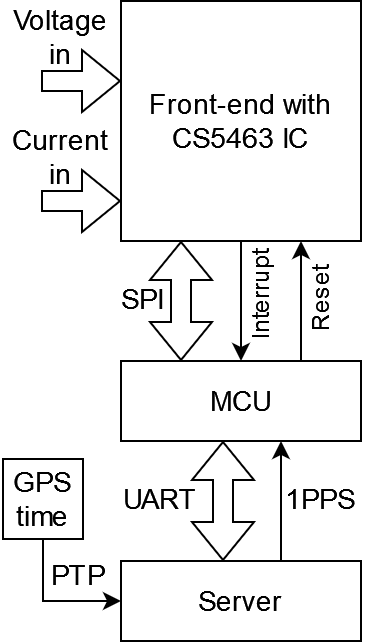}}
\caption{A block diagram overview of the platform.}
\label{general_diagram}
\end{figure}
\subsection{Front-end and energy measurement IC}
The front-end used in this proof-of-concept (PoC) prototype is a Cirrus Logic CDB5463U-Z evaluation board with an added signal conditioning circuit in the form of a voltage divider connected to the VIN+ input of the board (connector J23 \cite{CDB_datasheet}). The board, visible in Fig. 2, has separate differential voltage and current inputs. It already includes common-mode and differential noise RC filters on both inputs (components C9, C17 and C13 \cite{CDB_datasheet}), but the added voltage divider is necessary to limit the input signal amplitude to 500 mV\textsubscript{pp}, the maximum differential input amplitude which the included CS5463 energy measurement IC supports. To conserve the symmetry of the common-mode filters and prevent common-mode noise being converted into differential noise, an additional resistor of similar value to the total parallel resistance in the voltage divider was added to the VIN- input, while the evaluation board's filters are already configured such that the differential cutoff frequency is much lower than the common-mode cutoff frequency as well, thus further reducing the impact of possible asymmetry \cite{analog_ref}.
\begin{figure}[htbp]
\centerline{\includegraphics[scale=0.33]{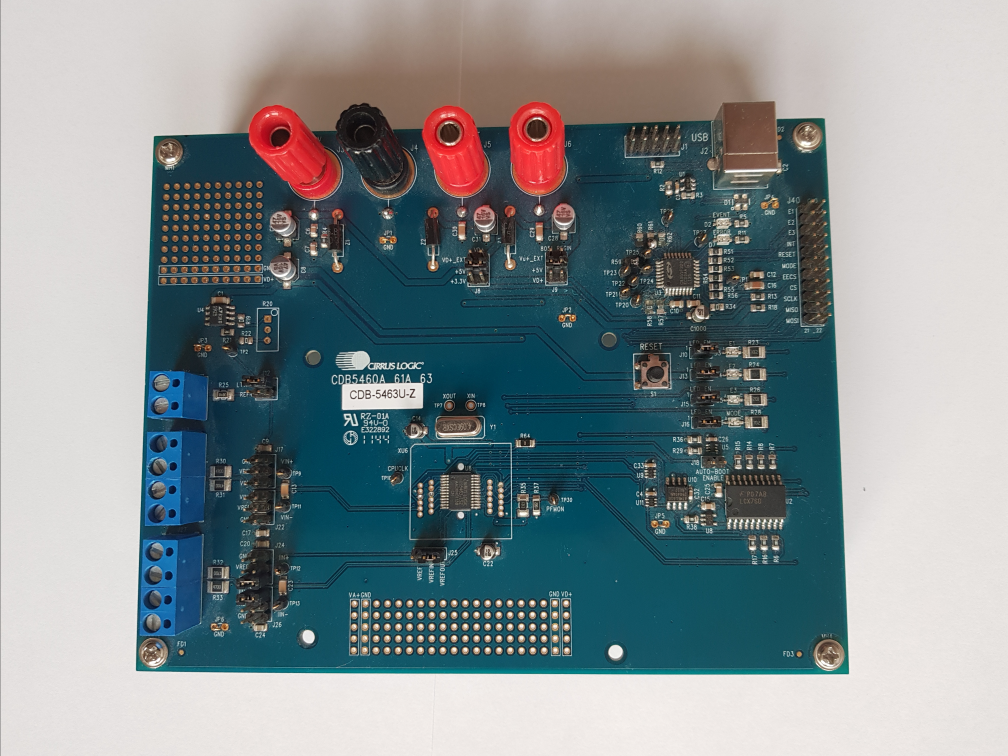}}
\caption{The CDB5463U-Z evaluation board.}
\label{triangle_wave_graph}
\end{figure}
\subsection{Microcontroller}
An MCU controls the CS5463 energy measurement IC, receives measurements from it and timestamps each measured value. In this PoC the MCU used is an STMicroelectronics STM32L072CZ on the B-L072Z-LRWAN1 development board. It communicates with the CS5463 using SPI (Serial Peripheral Interface) and additional connections for the Interrupt and Reset pin, while the connection to the server is made using the UART (Universal Asynchronous Receiver-Transmitter) protocol and a UART to USB bridge.

The MCU runs a program which receives accurate time data from the server, reads data (measurements) from the CS5463, timestamps them and sends them to the server. The program has several interrupt service routines (ISRs) handling UART communications and accurate timekeeping. The flow diagram is shown in Fig. 3.
\begin{figure}[htbp]
\centerline{\includegraphics[scale=0.27]{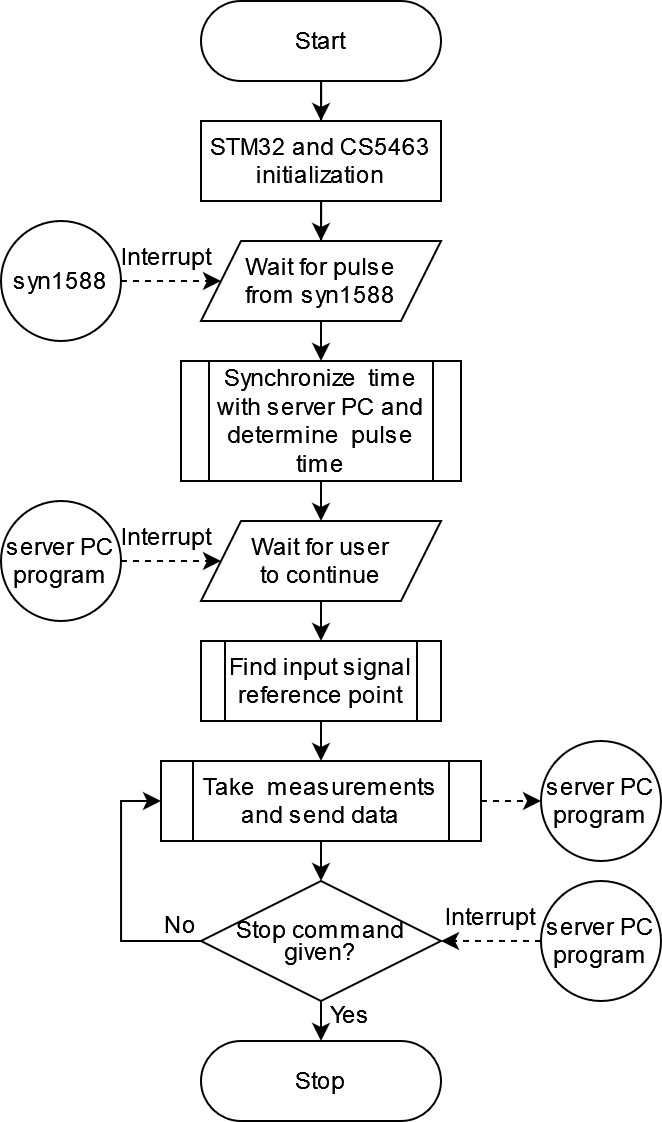}}
\caption{MCU program flow diagram.}
\label{mcu_flow_diagram}
\end{figure}

\subsubsection{Time synchronization with the server}
The synchronization process follows the standard Precision Time Protocol (PTP) synchronization algorithm \cite{PTP_protocol}. PTP message reception is done using an Oregano Systems syn1588 Network Interface Card (NIC, see section II-C). After starting the process, the MCU waits for a Sync message from the server which includes the current server time \texttt{t1}. When received, the MCU extracts the server time \texttt{t1} and takes the current MCU time \texttt{t2}. It then sends a Delay request message to the server, taking the sending time \texttt{t3} as well, and waits for a Delay response message. The Delay response message includes the server's Delay request receipt time \texttt{t4}. After receiving the Delay response message, the MCU has all 4 times stored and calculates the offset using the formula
\begin{equation}
    offset = ((t2 - t1) - (t4 - t3))/2
\end{equation}
and subtracts the offset from the current MCU time. As the process starts at the time of pulse reception from the syn1588 NIC, it takes the MCU time of synchronization start and MCU time of synchronization end. Using these values, it calculates the pulse reception MCU time \texttt{pulse\_time}.

\subsubsection{Timekeeping and interrupt service routines}
The resolution of time kept by the MCU is 0.1 ms. An ISR handles timekeeping using the \texttt{pulse\_time} variable. After the first synchronization, the MCU no longer requests actual timestamps from the server and only increases the \texttt{pulse\_time} variable by 1000.0 ms at each subsequent pulse, then sets MCU time to that time. Another ISR is responsible for increasing the internal timer by 1 every 0.1 ms, handling the interrupt created by the MCU's internal timer.
Stopping measurements is done using the \textit{user button} \cite{LRWAN_manual} on the development board. When the user presses this button, the program takes a final measurement, indicates to the server that a stop has been requested and stops.

\subsubsection{Data frame structure}
The data frames sent from the MCU subsystem to the server can be completely customized according to the desired purpose. For the purpose of this paper, determining accuracy of the platform, data frames consist of 12 bytes shown in Fig. 4. In this paper only one, voltage channel, is analyzed. The first byte is an indicator byte, indicating to the server that data frames are being sent. The next 6 bytes are measured data, the first 3 being the instantaneous voltage value and the last 3 the RMS voltage value. Measured data is transmitted unchanged with the server being responsible for parsing. Instantaneous voltage values are represented using normalized 2's complement, while RMS voltage values are represented in unsigned binary notation \cite{CS5463_datasheet}. The final 5 bytes are reserved for the timestamp. To reduce the amount of data being sent to the server and increase measurement speed, only a partial timestamp is kept and sent: a UNIX timestamp in microseconds divided by 100 (reflecting the resolution of time kept by the MCU, resulting in no data loss) and reduced by a constant (\num{16e12} at the time of writing).
\begin{figure}[htbp]
\centerline{\includegraphics[scale=0.16]{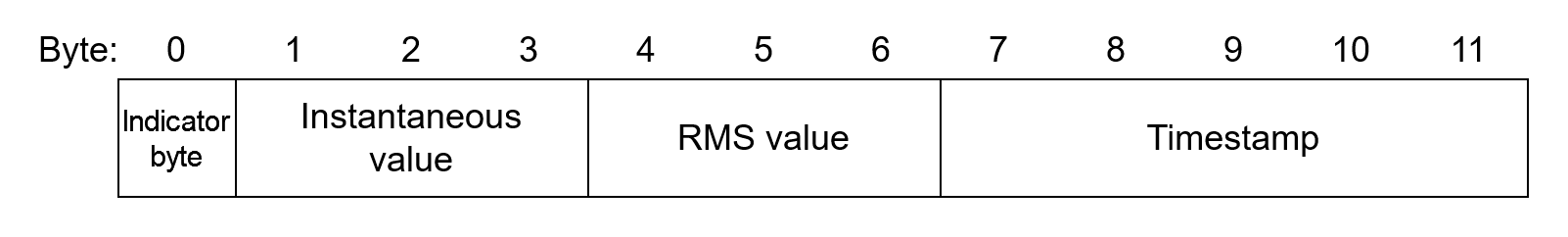}}
\caption{Data frame structure.}
\label{data_frame}
\end{figure}

\subsection{Server}
The server is a general-purpose PC running Microsoft Windows 10 with a multi-core CPU and an Oregano Systems syn1588 PCIe NIC. The syn1588 receives accurate GPS time through a PTP network and provides it through its API, as well as having a precise 1 pulse per second (1PPS) TTL-compatible output \cite{syn1588_datasheet}. The server runs a program which sets accurate time on the MCU subsystem, receives the timestamped measurements from it and saves them to a file. To decouple saving data, which can be a slow operation, from receiving data, the program is split into two threads. The program works as follows:
\begin{enumerate}
    \item After starting, the program elevates itself to the real-time Windows scheduling priority to reduce latency as much as possible, initializes the USB COM port and syn1588 NIC, and creates a file for writing data to. It also creates two threads which execute in parallel.
    \item The first thread is used for communication with the MCU subsystem. After the previous step, it takes an accurate timestamp from the syn1588 NIC using its API and synchronizes the time on the MCU using the PTP time synchronization algorithm described in section II-B. The program keeps track of the aforementioned constant and can increase it in order to not overflow the 5 bytes reserved for the timestamp. The thread then starts listening on the USB COM port for incoming data frames, which are read and passed to the second thread. If the user stops the measurement in the way described in section II-B, the thread will no longer accept data frames and will terminate.
    \item The second thread is used for saving data. It receives data frames from the first thread through a buffer and writes it to the created file. After no more data frames are received from the first thread when the user stops measurements, it will terminate.
    \item Finally, the program closes the USB COM port and the file and terminates.
\end{enumerate}

Data is saved in its raw binary format and must be parsed. This is done using a separate program which reads the raw data frame by frame, converts it into a readable format according to the frame description in section II-B and saves the converted measurements into a separate file.

\section{Measurements and results}
The platform's voltage input was connected directly to a Rigol DG4062 waveform generator as the signal source. A SEL-735 Power Quality and Revenue Meter that also includes PMU functions, was connected to the same source as well, in order to take measurements and determine accuracy of the described platform with respect to the PMU as the reference. The CS5463 was set to take 4000 instantaneous measurements per second with an RMS computation cycle of 80 measurements, resulting in a computation cycle frequency of 50 Hz, while the data frame was set according to the description in section II-B. Two sets of voltage measurements were taken, one of a sine and one of a triangle wave, both of \(f = 50 Hz\) frequency and \(V_{in} = 20 V_{pp}\) amplitude nominally. Each set includes about 10 minutes of measurements from both the platform and the reference PMU. A PMU 735 was connected to the local phasor data concentrator. The whole setup is synced to the UTC time using a PTP transparent Ethernet network. All measurements were done with the PMU and the platform at the same time in order to be directly comparable, which is possible due to all measurements being timestamped. From these large datasets, random measurements were taken and compared between the platform and the PMU. The comparison was done between differing numbers of measurements to assess the impact of comparison set size on calculated accuracy. For each comparison set, a mean value $\overline{V}$ and standard error $s$ were calculated for both the PMU and the platform using the formulas 
\begin{equation}
    \overline{V}=\frac{1}{n}\Sigma_{i=1}^{n}V_i
\end{equation}
\begin{equation}
    s=\sqrt{\frac{1}{n-1}\Sigma_{i=1}^{n}(V_i-\overline{V})^2}
\end{equation}
Additionally, a mean percentage difference was calculated using the formula
\begin{equation}
    |\overline{V}_{PMU}-\overline{V}_{platform}|/\overline{V}_{PMU}
\end{equation}
The voltage divider's voltage ratio was determined to be
\begin{equation}
    R_2/(R_1+R_2) = 0.02120
\end{equation}
using the calibrated Agilent 3458A high-precision multimeter.
\subsection{Sine wave}
Basic error comparison is provided. Results for comparison set sizes of n = 10, 20 and 30 measurements are shown in Table I and Fig. 5. The mean percentage difference is small (less than 0.1\%) but relatively consistent.

\subsection{Triangle wave}
Results for comparison set sizes of n = 10, 20 and 30 measurements are shown in Table II and Fig. 6. The mean percentage difference is again consistent but significantly larger than for the sine wave input.
\begin{table}[htbp]
\caption{Results for sine wave input}
\begin{center}
\begin{tabular}{|c|c|c|c||c|}
\hline
\multicolumn{2}{|c|}{}&{\textbf{PMU}}&{\textbf{Platform}}&{\textbf{Mean diff.}} \\
\hline
\multirow{2}{*}{n = 10}& $\overline{V}$/V & 6.9798 & 6.9829 & 0.04441\% \\
\cline{2-5}
& $s$/V & \num{1.0711e-3} & \num{6.2583e-4} & - \\
\hline
\multirow{2}{*}{n = 20}& $\overline{V}$/V & 6.9800 & 6.9829 & 0.04155\% \\
\cline{2-5}
& $s$/V & \num{1.1885e-3} & \num{7.2190e-4} & - \\
\hline
\multirow{2}{*}{n = 30}& $\overline{V}$/V & 6.9798 & 6.9829 & 0.04441\% \\
\cline{2-5}
& $s$/V & \num{1.4580e-3} & \num{1.0321e-3} & - \\
\hline
\end{tabular}
\label{sine_table}
\vspace{-20pt}
\end{center}
\end{table}

\begin{table}[htbp]
\caption{Results for triangle wave input}
\begin{center}
\begin{tabular}{|c|c|c|c||c|}
\hline
\multicolumn{2}{|c|}{}&{\textbf{PMU}}&{\textbf{Platform}}&{\textbf{Mean diff.}} \\
\hline
\multirow{2}{*}{n = 10}& $\overline{V}$/V & 5.6594 & 5.7006 & 0.7280\% \\
\cline{2-5}
& $s$/V & \num{1.0711e-3} & \num{6.2583e-4} & - \\
\hline
\multirow{2}{*}{n = 20}& $\overline{V}$/V & 5.6594 & 5.7007 & 0.7298\% \\
\cline{2-5}
& $s$/V & \num{1.1885e-3} & \num{7.2190e-4} & - \\
\hline
\multirow{2}{*}{n = 30}& $\overline{V}$/V & 5.6592 & 5.7006 & 0.7316\% \\
\cline{2-5}
& $s$/V & \num{1.4580e-3} & \num{1.0321e-3} & - \\
\hline
\end{tabular}
\label{triangle_table}
\end{center}
\end{table}

\begin{figure}[htbp]
\centerline{\includegraphics[scale=0.495]{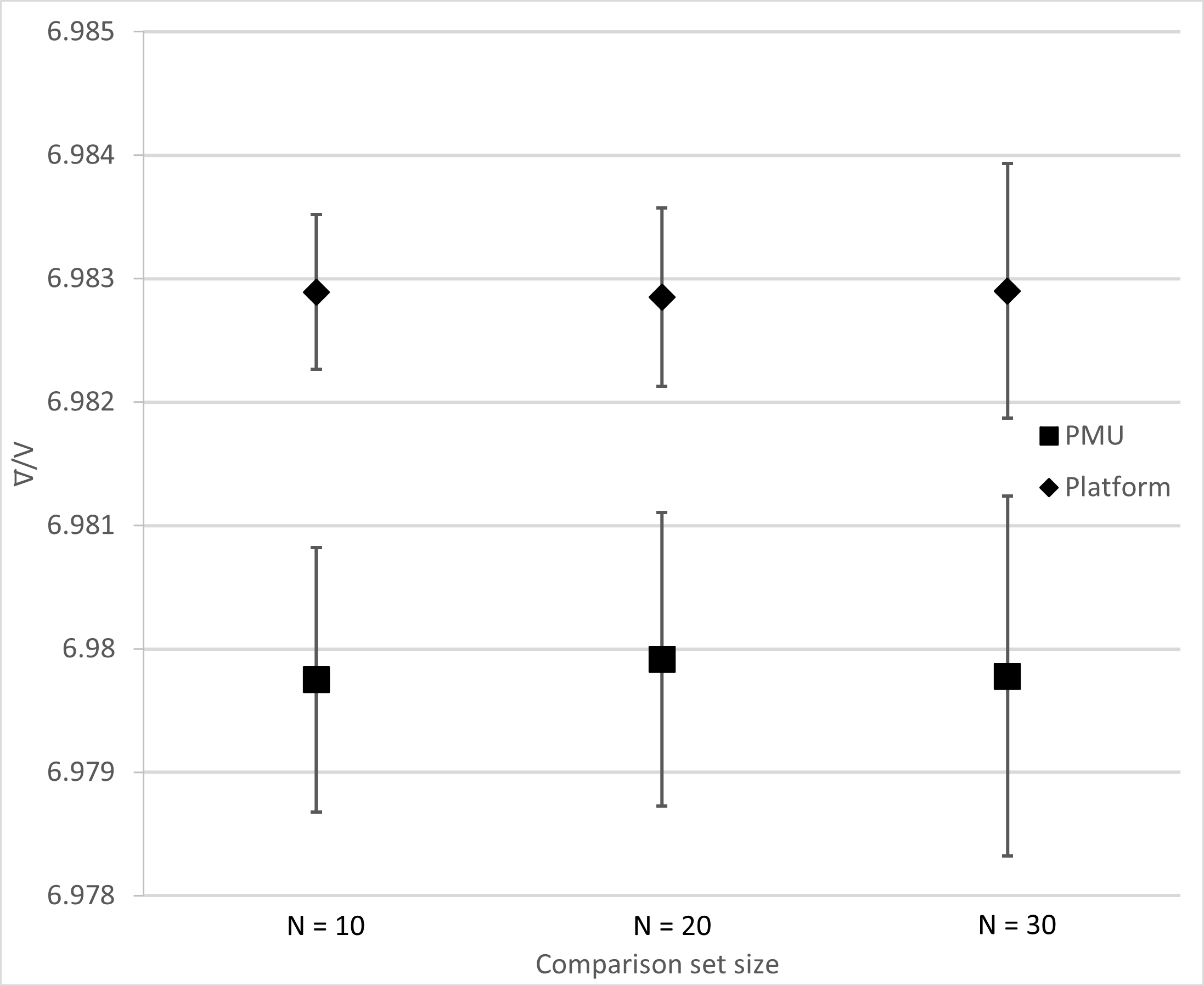}}
\caption{Sine wave input results.}
\label{sine_wave_graph}
\end{figure}

\begin{figure}[htbp]
\centerline{\includegraphics[scale=0.495]{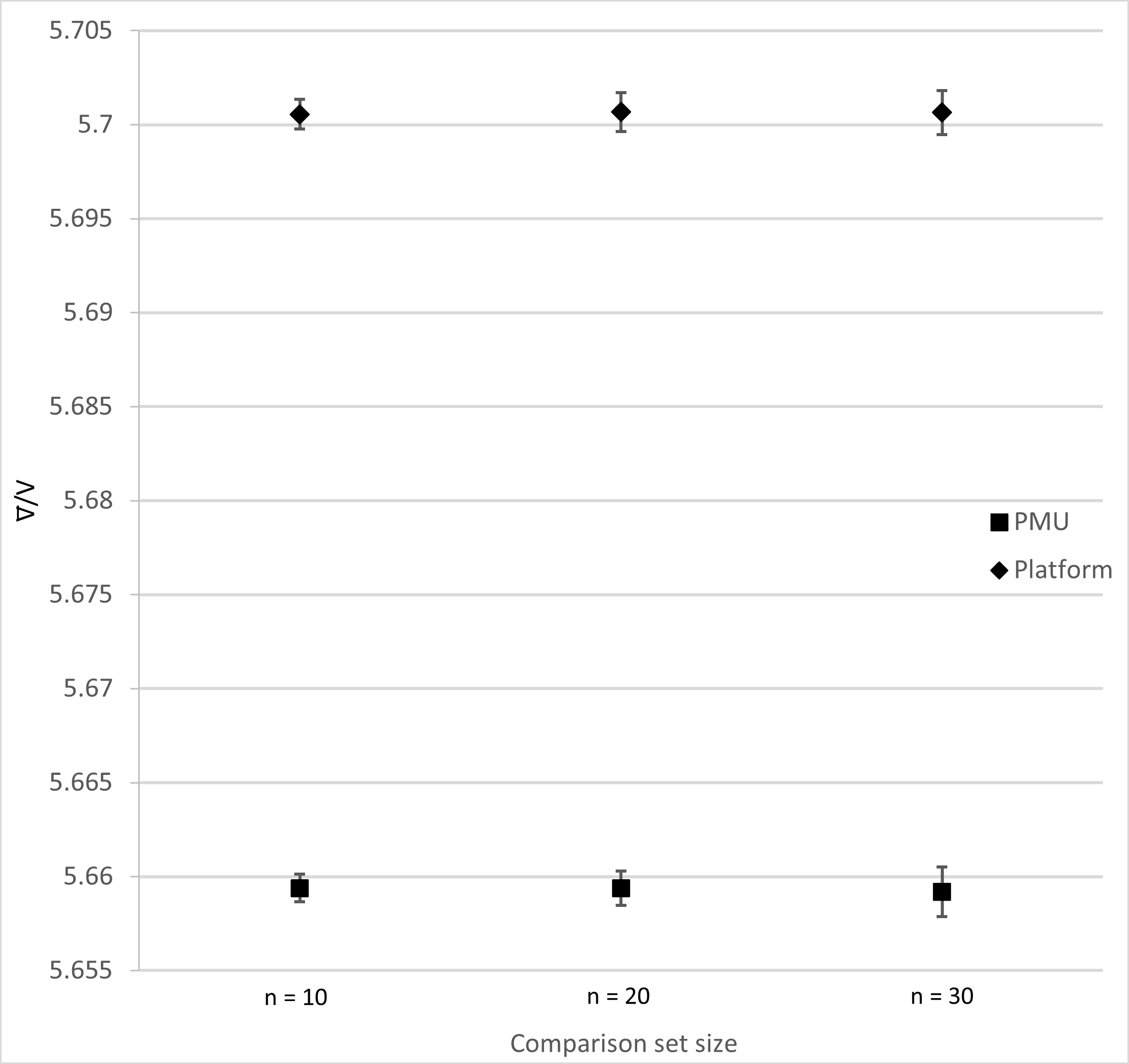}}
\caption{Triangle wave input results.}
\label{triangle_wave_graph}
\end{figure}

\subsection{Discussion}
Taking the 735 PMU as an accurate (calibrated) standard, the mean percentage difference calculated in sections III-A and III-B is a result of multiple possible factors which affect accuracy, particularly the CS5463's calibration, front-end frequency response and the possibly differing RMS calculation methods between the CS5463 and the PMU used. Since the standard error of the platform is small, indicating high precision, the mean percentage difference being consistent for both inputs is possibly due to an inaccuracy of calibration. However, for the triangle wave input the difference is significantly larger, although again consistent, with the standard error being small as well. In addition to calibration inaccuracies, since the triangle have contains a large amount of harmonics, the front-end frequency response and RMS calculation methods could have a much larger impact on measurements.

\section{Conclusion}
The paper describes a flexible educational platform with implemented basic PMU-like functionality. The platform has instantaneous and RMS voltage and current measurement capabilities and accurate time synchronization through PTP. Its accuracy was tested against a SEL-735 PMU as a standard and was found to be dependent on the input signal but overall a difference of less than 1\% was observed. The main reasons for development of the platform were low cost (relative to a standard PMU) and high flexibility and customizability for the user's needs, making it functional as an educational platform.

With further development of this proof-of-concept, similar functionality could be implemented in regular smart meters for monitoring of the distribution grid. Even though the sampling rate of all-in-one electricity metering IC solutions like the CS5463 is relatively low, it is sufficient for monitoring slow and medium-speed faults and events like slow voltage variations, voltage dips, interruptions and breaks, and low-frequency transients \cite{faults}. Specific improvements and challenges to overcome include:
\begin{itemize}
    \item Accurate timekeeping and simplification of the server. A smart meter cannot be connected directly to a PC with a NIC like the syn1588 used in this platform, but time of a large group of meters could be synchronized with an accurate time source through power line communication \cite{PLC}. Another possible simplification is using a more specialized and smaller server computer like the Raspberry Pi Compute Module 4 which includes integrated hardware PTP support \cite{RPICM4_PTP}. Being used as the server, such a device could provide better near-real time performance, lower power consumption, smaller size and even more system flexibility.
    \item High speed MCU. An MCU with a higher processor clock speed can provide a better time resolution (due to faster interrupts and more precise clock divider) as well as a higher sampling rate from the metering IC. Another option is a custom-built IC or an FPGA.
    \item Metering IC with a higher sampling rate or custom-built ADC and DSP.
    \item Better front-end based on a transformer or Hall sensor for higher accuracy.
\end{itemize}

\end{document}